# One dimension spring supported ball on top of a sinusoidal vibrating plate: A forced oscillation simulation using molecular dynamics method

Sparisoma Viridi[1,*], Wahyu Srigutomo[2], Wahyu Hidayat[3], Alamta Singarimbun[2], and Sitti Balkis[4]

[1]Nuclear Physics and Biophysics Research Division, Insitut Teknologi Bandung, Indonesia
[2]Physics of Complex System Research Division, Insitut Teknologi Bandung, Indonesia
[3]Theoretical and High Energy Physics Research Division, Insitut Teknologi Bandung, Indonesia
[4]Magister Program in Physics Teaching, Institut Teknologi Bandung, Indonesia
[*]dudung@fi.itb.ac.id

Absctract

A ball supported by a spring is set on top of a plate which is sinusoidal vibrated. The motion is limited to one dimension motion. It is assumed that the spring is an ideal one with zero mass. The vibrating plate is considered much heavier than the ball, so that the ball motion has no influence on the plate motion. Plate vibration frequency is varied around the frequency of ball-spring system. Resonance phenomenon is reported, which needs a phase match condition to occur.

## Introduction

Forced oscillation is a well known type of oscillation system beside the damped and ordinary oscillation. It has many applications as model in our daily life, such as in chemical reaction [1, 2], electronic circuits [3], and mechanical vibration [4, 5]. Even it is already an established topic, there still a chance to implement it on other system. Usual parametes to be varied are forcing frequency and amplitude [6]. Recent experiments [7, 8] need an explanation, which can address to series of forced oscillation that forms a speaker membrane.

## Forced oscillation system

A ball, a spring, and a sinusoidal vibrating plate construct a forced oscillation system as illustrated in Figure 1. Upward direction is taken to be in positive $x$-direction, while the earth gravitation acceleration $g$ is set in the other direction. Motion of the ball is considered to be only in one dimension.

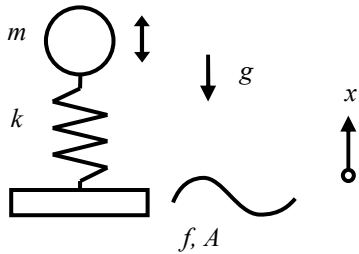

Figure 1. Forced oscillation system: a ball with mass $m$, a spring with spring constant $k$, and a sinusoidal vibrating plate with frequency $f$ and amplitude $A$.

The plate is much heavier than the ball. It means that the oscillation motion of the ball does not affect the motion of the plate. It simplifies the system that only the ball and the spring which are to be considered. The spring force depends on the distance between the ball and the plate, and also to the initial length of the spring $l_0$.

Position of the plate is defined as $x_p(t)$, which is actually function of the forcing oscillation

$$x_p(t) = x_{p0} + A\sin(2\pi f t + \phi_0) . \qquad (1)$$

Then spring force $F_k$, which is suffered by the ball, defined by

$$F_k = -k\left[x(t) - x_p(t) - l_0\right], \qquad (2)$$

where $x(t)$ is position of the ball. Other considered forces are earth gravitational force $F_g$ dan air friction force $F_f$, which are

$$F_g = -mg , \qquad (3)$$

and

$$F_f = -3\pi\eta D \frac{dx}{dt} , \qquad (4)$$

where $D$ is ball diameter and $\eta$ is air viscosity. Using Newton's second law of motion with Equation (1)-(4) a differential equation in a form of

$$\frac{d^2x}{dt^2} + c_1\frac{dx}{dt} + c_2 x + c_3 = c_4 \sin(2\pi f t + \phi_0) \qquad (5)$$

can be obtained, with

$$c_1 = \frac{3\pi\eta D}{m} , \qquad (6)$$

$$c_2 = \frac{k}{m} , \qquad (7)$$

$$c_3 = \left[g - c_2(x_{p0} + l_0)\right], \qquad (8)$$

$$c_4 = c_2 A . \qquad (9)$$

Equation (5), with the definitions from Equation (6)-(9), will be solved numerically using molecular dynamics (MD) method implementing Gear predictor-corrector (GPC) algorithm of 5th order [9].

## Gear predictor-corrector algorithm

The algoritm has two steps. The first step is formulated as follow

$$\begin{pmatrix} \vec{r}_0^{\,p}(t+\Delta t) \\ \vec{r}_1^{\,p}(t+\Delta t) \\ \vec{r}_2^{\,p}(t+\Delta t) \\ \vec{r}_3^{\,p}(t+\Delta t) \\ \vec{r}_4^{\,p}(t+\Delta t) \\ \vec{r}_5^{\,p}(t+\Delta t) \end{pmatrix} = \begin{pmatrix} 1 & 1 & 1 & 1 & 1 & 1 \\ 0 & 1 & 2 & 3 & 4 & 5 \\ 0 & 0 & 1 & 3 & 6 & 10 \\ 0 & 0 & 0 & 1 & 4 & 10 \\ 0 & 0 & 0 & 0 & 1 & 5 \\ 0 & 0 & 0 & 0 & 0 & 1 \end{pmatrix} \begin{pmatrix} \vec{r}_0(t) \\ \vec{r}_1(t) \\ \vec{r}_2(t) \\ \vec{r}_3(t) \\ \vec{r}_4(t) \\ \vec{r}_5(t) \end{pmatrix} \quad (10)$$

and the second step is as

$$\begin{pmatrix} \vec{r}_0(t+\Delta t) \\ \vec{r}_1(t+\Delta t) \\ \vec{r}_2(t+\Delta t) \\ \vec{r}_3(t+\Delta t) \\ \vec{r}_4(t+\Delta t) \\ \vec{r}_5(t+\Delta t) \end{pmatrix} = \begin{pmatrix} \vec{r}_0^{\,p}(t+\Delta t) \\ \vec{r}_1^{\,p}(t+\Delta t) \\ \vec{r}_2^{\,p}(t+\Delta t) \\ \vec{r}_3^{\,p}(t+\Delta t) \\ \vec{r}_4^{\,p}(t+\Delta t) \\ \vec{r}_5^{\,p}(t+\Delta t) \end{pmatrix} + \begin{pmatrix} c_0 \\ c_1 \\ c_2 \\ c_3 \\ c_4 \\ c_5 \end{pmatrix} \Delta \vec{r}_2(t+\Delta t) . \quad (11)$$

The term $\Delta \vec{r}_2(t+\Delta t)$ is defined as

$$\Delta \vec{r}_2(t+\Delta t) = \vec{r}_2(t+\Delta t) - \vec{r}_2^{\,p}(t+\Delta t) . \quad (12)$$

Upper index $p$ indicates the prediction step (first step), while the correction step (second step) has no index, since it already gives the corrected value of all motion parameters. Motion parameters are defined as

$$\vec{r}_n(t) = \frac{(\Delta t)^n}{n!}\left[\frac{d^n \vec{r}_0(t)}{dt^n}\right], \quad (13)$$

which have dimension of distance.

## Results and discussion

In this work the resonance frequency $f_0$ is chosen to have value 20 Hz. Mass of the ball is set to 0.1 kg, then the spring constant should be about 1579 N·m. Then the oscillation period $T_0$ would be 0.05 s. It is usual to take the time discretization $\Delta t$ to be hundred times smaller than an important physical time [9]. In this case a full period of a sinusoidal cycle is divided into 50 points then $\Delta t$ is set to 1/100 to this value, which gives $\Delta t = 10^{-5}$ s. $A = 0.01$ m and $l_0 = 0.01$ m are used in the simulation. For air viscosity, the value is taken for room temperature 25 °C using Sutherland's formula [10], which gives value of 18.616 µPa·s. Mentioned values lead to

$$c_1 = 3.509 \times 10^{-5},$$

$$c_2 = 1.579 \times 10^3,$$

$$c_3 = -6.850 \times 10^1,$$

$$c_4 = 1.579 \times 10^1.$$

The influence of initial phase $\phi_0$ can be seen in Figure 1, that different value of it gives different pattern in position of the ball. The patterns are unique in the range $\phi_0 \in [0, \frac{1}{4}\pi]$, while in other range they are only vertically or horizontally reflected patterns to the patterns in the range.

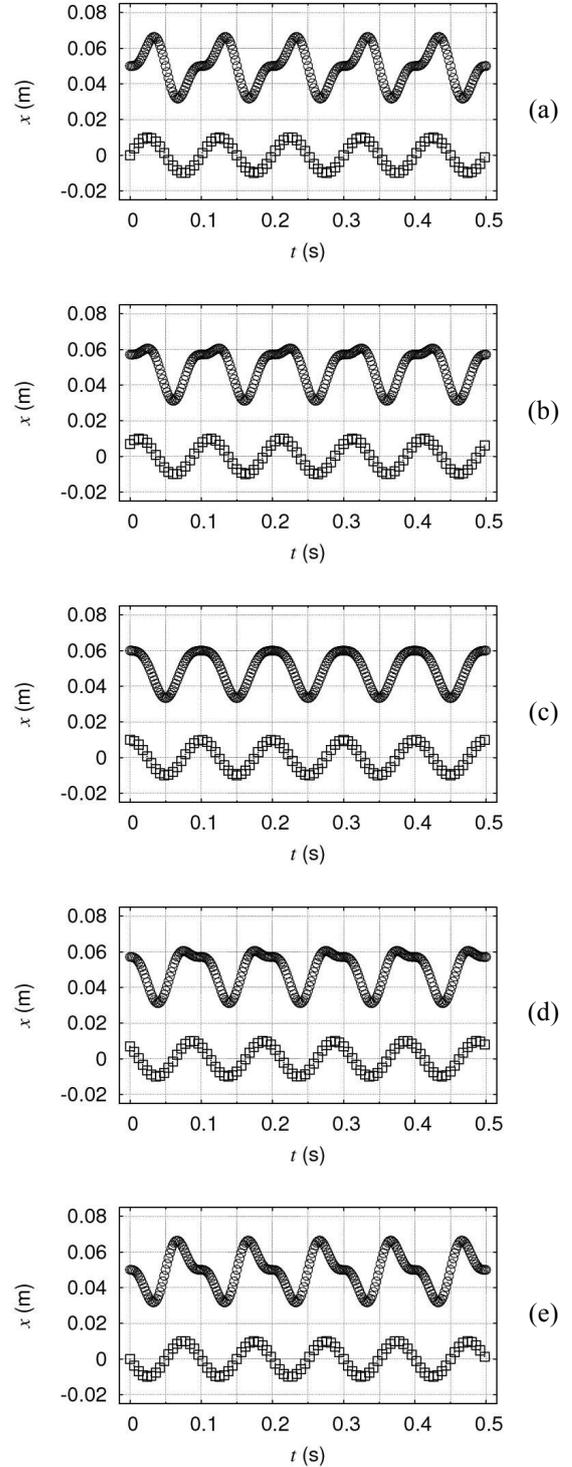

Figure 1. Typical motion of spring attached ball (o) and vibrating plate (□) for vibrating fre-quency $f = 10$ Hz, vibrating amplitude $A = 0.01$ m, resonance frequency $f_0 = 20$ Hz, and: (a) $\phi_0 = 0$, (b) $\phi_0 = \frac{1}{4}\pi$, (c) $\phi_0 = \frac{1}{2}\pi$, (d) $\phi_0 = \frac{3}{4}\pi$, and (e) $\phi_0 = \pi$.

A parameter similar to two times amplitude $2A$ is defined as distance peak to peak of the $x(t)$ of the ball. It can be seen in Figure 2 that this parameter has minum dan maximum value at $\phi_0$ about $\phi_0 = n\pi$ and $\phi_0 = (n+\tfrac{1}{2})\pi$, with $n = 0, 1, 2, 3, ..$, respectively.

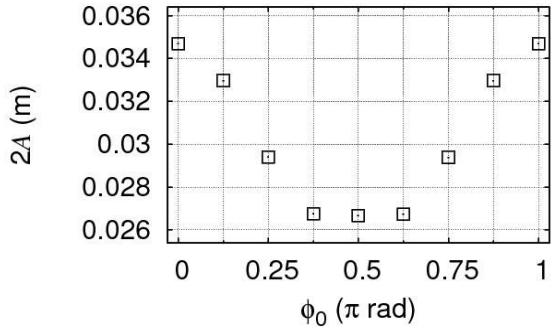

Figure 2. Peak to peak distance $2A$ as function of initial phase $\phi_0$ for $f = 10$ Hz, $A = 0.01$ m, and $f_0 = 20$ Hz.

For investigating influence of the frequency $f$ value of $\phi_0 = 0$ is chosen and the result is as shown in Figure 3.

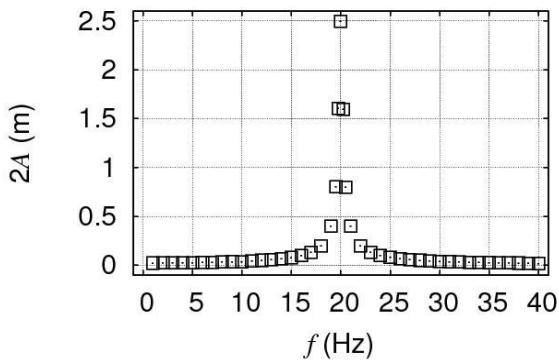

Figure 3. Peak to peak distance $2A$ as function vibration frequency $f$ with resonance frequency $f_0 = 20$ Hz.

As it is expected there is a peak of $2A$, which is coincide at the value $f = f_0$ as an usual forced oscillation system, even though the time dependent external force, left side of Equation (5), does not work directly to the ball but through the spring. But it produces similar resonance effect as the usual system.

It is also interesting to investigate how strong the initial phase $\phi_0$ influencing the peak to peak $2A$ value at the same time with vibration frequency $f$. The results for this calculation is given in Figure 4. Obviously, the role of $\phi_0$ is not as strong as $f$. This means that in order to achieve a resonance state, the vibration frequency $f$ is more important than the initial phase $\phi_0$.

If $2A$ at $f = 10$ Hz is chosen as reference value, then the influence of $\phi_0$ gives only about 20 % amplification, while the influence of $f$ gives about 1000 % amplification.

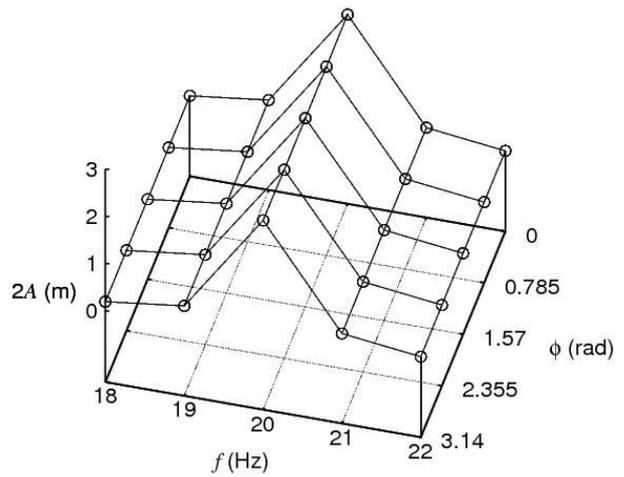

Figure 4. Peak to peak distance $2A$ surface as function of vibration frequency $f$ and initial phase $\phi_0$, where resonance frequency $f_0$ is about 20 Hz.

**Conclusion**

A simulation of a spring attached ball that is put on top of a sinosoidal vibrating plate has been performed. Even the vibrating plate does not touch the ball directly but trough the spring, the final equation and the simulation results confirm that the system is not than a forced oscillation system, which also exhibits resonance phenomenon. Initial phase of plate vibration do has influence on the amplification of peak to peak distance of the ball oscillation, about 20 %, while the vibration frequency about 1000 % compare to the value at 10 Hz vibration frequency.

**Acknowledgement**

Authors would like to thank to Institut Teknologi Bandung Alumni Association (IA-ITB) Research Grant in 2011 for supporting simulation part of this work. Preparation of this manuscript is supported by Indonesian Journal of Physics (IJP) Internationalization Program in 2010 financed by Higher Education Directorate (Dikti) of Republic of Indonesia.